\documentclass[a4paper,10pt]{article}
\usepackage[utf8]{inputenc}
\title{de Sitter Spacetime as a Natural Superconductor}
\author{Davood Momeni$^1$, Ratbay Myrzakulov$^1$, Zaid Zaz$^2$\\
\\ $^1$Eurasian International Center for Theoretical Physics\\ and
Department of General Theoretical Physics,\\ Eurasian National
University, Astana 010008, Kazakhstan\\ 
\\$^2$Department of Electronics and Communication Engineering, \\University of Kashmir, 
Srinagar, Kashmir-190006, India}
\date{}
\begin{document}
\maketitle
\begin{abstract}
Motivated by the studies done on magnetically induced superconduc-
tivity on QCD vaccum we propose that de Sitter spacetime is a natural
superconductor. This is due to the occurrence of spinor condensation. We
provide a framework for joining the curvature of spacetime with super-
conductor processes. We demonstrate that for a critical value of de Sitter
radius unstable tachyon modes exist showing the existence of a supercon-
ducting phase
\end{abstract}

It has been seen that in presence of a strong magnetic field the QCD vacuum undergoes a phase transition where charged $\rho^{\frac{+}{-}}$ mesons are condensed. As a result the vacuum behaves as an in-homogeneous superconductor. In fact this induced superconductivity is an-isotropic and is a consequence of a non-minimal coupling of the $\rho^{\frac{+}{-}}$ mesons to the electromagnetic field \cite{1}. It has been observed that the superconducting ground state is like an  Abrikosov lattice state in an ordinary type-II superconductor. In fact, it is an inhomogeneous structure composed of a quark-anti quark condensate pierced by vortices. For such vortices acoustic vibrational modes have been studied and they are seen to possess a linear (quadratic) dispersion relation corresponding to type I (type II) Nambu-Goldstone modes \cite{2}. Magnetically induced superconductivity has also been seen in the electroweak quantum vaccum. In fact a super-fluid behavior is also induced due to the condensation of W and Z bosons. It is seen that this  superconductor-super fluid phase transition occurs at the critical magnetic field of $10^{20}T$ \cite{3}.  Magnetically induced superconductivity in QCD vaccum has also been investigated from an effective low energy model of QCD known as Nambu-Jona-Lasinio model. It is seen that the superconducting vacuum is composed of a new type of vortices. These vortices are topological defects in the charged vector condensates. The superconductivity is present only along the axis of the magnetic field. It is seen that this effect is absent in QED \cite{4}. 

Transition to type-II superconducting phases are also seen in the core of cooling neutron stars. It is seen that the transport of magnetic flux inside the liquid core of a neutron star is due to the interaction between neutron and proton super-fluid vortices during intervals of spin-down or spin-up in binary systems \cite{5}. However it has been pointed out that the standard picture of the neutron star core being made up of a mixture of a neutron super-fluid and a proton type-II superconductor is not consistent with observations of a long period precession in isolated pulsars. In fact the vortex-vortex interaction is modified and the resulting superconductor is a type-I superconductor. It may be noted that the magnetic field is expelled from the superconducting regions of the neutron star which leads to the formation of the intermediate state when alternating domains of superconducting matter and normal matter coexist \cite{6}. 

Recent investigations have also revealed that many properties of strongly coupled superconductors can be potentially described by classical general relativity living in one higher dimension, which are referred to as holographic superconductors \cite{7}. It may be noted that holographic superconductors are Type II, which means that they start in a normal phase at large magnetic fields and low temperatures, they develop superconducting droplets as B is reduced \cite{8}. Furthermore it is known that a holographic superconductor dual exists for Lifshitz black hole. In such dual superconductors it is seen that the critical temperature for condensation is affected by the Lifshitz exponent $z$ \cite{9}. Conductivity and entanglement entropy of such high dimensional holographic superconductors has also been studied. It is seen that with an increase in dimensionality, the entanglement entropy decreases, the coherence peak in the conductivity becomes narrower. Also ratio between the energy gap and the critical temperature decreases. These results imply that the condensate interactions become weaker in high spatial dimensions \cite{13}. Neutral and electrically charged black-holes have also been studied in a variety of limits. These have been shown to produce essential components of superconductivity in the dual $2+1$ dimensional field theory, forming a condensate below a critical temperature. It may be noted that this superconductor can be placed in an external magnetic field by adding magnetic charges to the black hole. It is seen that a family of condensates may be formed. When the magnetic field is finite, these condensates are localized in one dimension. It may be noted that their profile is exactly solvable since it maps to the quantum harmonic oscillator. With increase in the magnetic field, the condensate shrinks in size, which is reminiscent of the Meissener effect \cite{10}. 

In this letter, using the magnetically induced superconductivity of the QCD vaccum as a motivation, we study spinor condensation in de Sitter spacetime and show that de Sitter spacetime is a natural superconductor.
In fact induced superconductivity in de Sitter spacetime may provide interesting insights into the infra-red cut-offs. It may be noted that history of the de Sitter spacetime plays an important role in resolving the infrared problems. It is seen the diagonalization of the Hamiltonian for long-wavelength modes leads to an infrared cutoff. It may be noted that this bears resemblance to a bosonic superconductor in which graviton-pairing occurs between non-adiabatic modes \cite{11}. Furthermore it is seen that in presence of a superconducting dark energy, the cosmological evolution of the universe ends in an exponentially accelerating vaccum de Sitter state \cite{12}.
de Sitter spacetime is defined to be a spacetime of positive constant curvature, de Sitter spacetime has a maximal degree of space and time symmetry and as a result a quantum field theory under  the group of special relativistic rotations and translations can be formulated in it. Even though formulating a co-variant quantum field theory in curved spacetime is difficult \cite{12a}. The topology of de Sitter spacetime is $\mathcal{R} \times \mathcal{S}^{3}$ and can be effectively viewed as a hyperboloid in five dimensional minkowski spacetime. The global coordinate form of the de Sitter metric can be obtained by  standard global compactification scheme of five dimensional pseudo-riemannian flat spacetime as follows.

$x^0=L\cos\theta_1\cosh\theta_4,  x^1=L\sin\theta_1\sin\theta_2\cos\theta_3,  x^2=L\sin\theta_1\sin\theta_2\sin\theta_3, x^3=L\sin\theta_1 \cos\theta_2,  x^4=L\cos\theta_1\sinh\theta_4.$  \cite{n}-\cite{f}.
Thus the de sitter metric can be written as;\\
\begin{eqnarray}
 ds^2=(dx^4)^2-\Sigma_{i=0}^{3}(dx^i)^2
\end{eqnarray}
in this space, the wave equation for an arbitrary spinor field $\psi$ with a rest mass $\mu$ and an arbitrary spin $s$ can be written as \cite{PTP};
\begin{eqnarray}
\Big(\frac{1}{2 L^2}J^{\mu\nu}J_{\mu\nu}-s\mu^2\mp\frac{3i\mu\sqrt{s}}{L}\Big)\Psi=0
\end{eqnarray}
where, $L$ is de Sitter radius, $J_{\mu\nu}=x_{\mu}\partial_{\nu}-x_{\nu}\partial_{\mu}$ \cite{14}.
Thus Eq.(2) can be conveniently solved in spherical polar coordinates. A separable solution in spherical polar coordinates can be written as, 
\begin{eqnarray}
\Psi=\Pi_{i=1}^{4}\psi_{i}(\theta_i)=\psi_1(\theta_1)e^{im\theta_3+i\omega \theta_4}P_{l}^{m}(\cos\theta_2)
\end{eqnarray}
where, $P_{l}^{m}$ are associated Legendre functions. To solve the differential equation satisfied by $\psi(\theta_{1})$, we set  $x=\cos 2\theta_1$. Subsequently the following exact solution is obtained.
\begin{eqnarray}
\psi_1(\theta_1)=(1-x)^{\alpha/2-1/4}(1+x)^{\beta/2} F(-n,n+\alpha+\beta+1,\alpha+1)
\end{eqnarray}

In order to fully determine  $\psi(\theta_{1})$, the following algebraic equations are solved, 
\begin{eqnarray}
&&\alpha=\pm(l+\frac{1}{2}),\ \ \beta=\pm\omega,\\&&
n(n\pm(l+\frac{1}{2})\pm i\omega+1)+\frac{1}{2}(1\pm(l+\frac{1}{2}))(1\pm i\omega)\\&&\nonumber=
\frac{1}{4}\Big(3-l(l+1)+\omega^2-L^2(s\mu^2\pm\frac{3i\mu\sqrt{s}}{L})\Big)
\end{eqnarray}
We interpret $\omega_{1,2}^{\pm}$ as the two modes of the spinor field. These modes can be explicitly written out as,
\begin{eqnarray}\label{omega}
\omega_{1,2}^{\pm}=\left[ \begin {array}{c} i \left( l+2\,n+3/2 \right) 
\\ {\medskip}i \left( l-2\,n-1/2 \right) \end {array} \right] \pm\Big(\sqrt{s}\mu L\pm\frac{3i}{2}\Big).
\end{eqnarray}
It may be noted that the previous setup consists of an array of four possible modes, and operates in the $10^{-3}$ spectral region. These modes are sensitive to change in quantum numbers $n$ and $l$. For  $n,l=0$ we obtain the ground state as,
\begin{eqnarray}
\omega_{1,2}^{(0)\pm}=\left[ \begin {array}{c} \frac{3i}{2}
\\ \noalign{\medskip}-\frac{i}{2}\end {array} \right] \pm\Big(\sqrt{s}\mu  L\pm\frac{3i}{2}\Big).
\end{eqnarray}
It can be seen that , the energy of $\omega_{1,2}^{(0)-}$ is  lower than for those having $\omega_{1,2}^{(0)+}$.
As the sign for  $||\omega_{1,2}^{(0)\pm}||^2$ is fixed by the second bracket, we have the following possibilities for each mode,\\
\begin{eqnarray}
\omega_{1,2}^{(0)+}= \left[ \begin {array}{c} 3\,i+\sqrt {s}\mu\,L\\ \noalign{\medskip}
\sqrt {s}\mu\,L\\ \noalign{\medskip}-\sqrt {s}\mu\,L
\\i-\sqrt {s}\mu\,L\end {array} \right] \nonumber
\end{eqnarray}
\begin{eqnarray}
\omega_{1,2}^{(0)-}=
\left[ \begin {array}{c} i+\sqrt {s}\mu\,L\\ \noalign{\medskip}-2\,i+
\sqrt {s}\mu\,L\\ \noalign{\medskip}-2\,i-\sqrt {s}\mu\,L
\\ \noalign{\medskip}i-\sqrt {s}\mu\,L\end {array}\right] \nonumber
\end{eqnarray}

Now allowing for the mass of the spinor to assume complex values using the following definition,  $M^{\pm}_{X}(\mu,L;s)=\omega_{1,2}^{(0)\pm}$. We observe that for the pair of real masses, mass increases linearly with respect to the de Sitter radius $L$. Therefore no condensation occurs. However for the complex valued mass, we observe a tachyonic spectrum. It may be noted that it is independent from the form of spinor and the basic characteristic properties of it like $s,\mu$ and even the scalar curvature.  It is adequate to define the Euclidean norm of $||\omega_{1,2}^{(0)\pm}||^2$ as mass $M^{(0)^{\pm}}_{X}(\mu,L;s)$ as follows:
\begin{eqnarray}
&&M_{+}^2=4s\mu^2 L^2-18,\\&&
 M_{-}^2=4s\mu^2 L^2-10.
\end{eqnarray}
Once de Sitter radius has become  $L\leq\frac{3}{\mu\sqrt{2s}}$ ,the mass term  $M_{+}^2$ becomes negative and the tachyon remains. We expect an instability and consequently a superconducting phase  of matter. Similarly, superconductivity exists in $M_{-}^2$ case also.

Thus we were able to demonstrate that for a critical value of the de sitter radius, de Sitter spacetime can be effectively viewed as a superconductor.
In this letter we used the magnetically induced superconductivity of QCD vaccum as a motivation to study spinor condensation in de Sitter spacetime. We analyzed the wave equation for an arbitrary spinor field in de Sitter spacetime and demonstrated that if we allow the masses to assume complex values,then for a critical de Sitter radius a techyonic spectrum is observed which indicates the existence of a superconducting phase.
It may be noted that de Sitter spacetime is an important aspect of study in inflationary cosmology  \cite{15}-\cite{15a}.  It would be interesting to analyze what effect can spinor condensation have on inflation.  It may be noted that the  current observations strongly suggest that our Universe is accelerating in its  expansion and it may approach de Sitter spacetime asymptotically \cite{16}.  So, it might be possible that the spinor condensation can effect this  asymptomatic state of the Universe. 
Finally, we would like to point out that different deformations of de Sitter have been studied. 
It has recently been seen that physics at short distance in an inflationary scenario in de Sitter spacetime is deformed by the generalized uncertainty principle \cite{17}. This deformation contains an arbitrary numerical parameter, whose value is fixed from experiments 
\cite{18}. It will be interesting to study the effect of this deformation of de Sitter on the results obtained in this paper.  
It may be noted that recently the gravity's rainbow has been used to study the deformation of various geometries 
\cite{black1}-\cite{black2}.  So,  it would be interesting to deform the de Sitter spacetime by rainbow functions, and then 
analyze what effect such a deformation will have on spinor condensation.  It may be noted    rainbow functions 
have been motivated from results   obtained 
in loop quantum gravity  \cite{AmelinoCamelia:1996pj} and the hard spectra from gamma-ray 
burster's~\cite{AmelinoCamelia:1997gz}. Thus, the   de Sitter can be deformed using both  functions, and both these functions are expected to have interesting effects on the spinor condensation. 

\section*{Acknowledgments}
We would like to thank Igor I. Smolyaninov and   Mir Faizal  for useful discussions/comments

\end{document}